# Capturing Data Uncertainty in High-Volume Stream Processing

Yanlei Diao[†], Boduo Li[†], Anna Liu[†], Liping Peng[†], Charles Sutton[§],
Thanh Tran[†] Michael Zink[†]
[†]University of Massachusetts Amherst    [§]University of California Berkeley

## ABSTRACT

We present the design and development of a data stream system that captures data uncertainty from data collection to query processing to final result generation. Our system focuses on data that is naturally modeled as continuous random variables such as many types of sensor data. To provide an end-to-end solution, our system employs probabilistic modeling and inference to generate uncertainty description for raw data, and then a suite of statistical techniques to capture changes of uncertainty as data propagates through query operators. To cope with high-volume streams, we explore advanced approximation techniques for both space and time efficiency. We are currently working with a group of scientists to evaluate our system using traces collected from real-world applications for hazardous weather monitoring and for object tracking and monitoring.

## 1. INTRODUCTION

As sensor networks continue to expand into various aspects of everyday life, we are witnessing uncertain data streams, where data is *incomplete*, *imprecise*, and even *misleading*, emanating from a variety of environments. For instance, uncertain data streams have been observed in environmental monitoring sensor networks [18], Radio Frequency Identification (RFID) networks [33, 61], GPS systems [35], camera sensor networks [39], and radar sensor networks [40]. As such data streams are fed into existing stream processing systems to support tracking and monitoring applications, the results presented to end applications are often of unknown quality. Furthermore, volumes of such data streams and performance requirements of real-world applications preclude the use of complex data analysis in an offline manner.

Our research is particularly motivated by two emerging sensing applications. The first is object tracking and monitoring using RFID technology, in particular, wide-range mobile readers that enable cost-effective deployments in inventory management [22], healthcare [22], library management [20], etc. Data streams from mobile RFID readers carry primitive information about sensed objects and are highly noisy due to the sensitivity of sensing to the orientation of reading and many environmental factors. When such streams are used to support monitoring applications, e.g., to detect safety violations regarding flammable objects, the quality of alerts sent to the end application is of significant concern.

The second application is hazardous weather monitoring using radar networks [40]. In this application, a wide-area radar network collects a large amount of meteorological data and feeds such data into a real-time processing system to predict natural disasters such as tornados and severe storms. Uncertainty in such data arises in radar scanning, data compression, merging of different radar streams, etc. Prediction results derived from such data can thus be error-prone. Given the potential social impact of this system, it is absolutely vital to capture the quality of its prediction results.

The two applications above present a number of challenges to data stream systems: (*i*) Observed data is inherently incomplete and noisy, and the noise varies with time and location. (*ii*) Observed data is different from data needed for further processing. In RFID-based tracking and monitoring, observed data contains tag ids of objects while data of interest to monitoring applications concerns object locations. In weather monitoring, observed data is raw signal data whereas data needed for further processing is a numeric description of each unit area of space in terms of reflectivity, wind speed, etc. Hence, given raw data streams, a stream system needs to handle both the mismatch between observed data and data of interest and the noise in observed data. (*iii*) In these applications, data streams can arrive at a rate higher than in traditional sensing applications, e.g., 200Mb/sec from a single radar node. Such high volumes require that processing of raw data must keep up with stream speed. (*iv*) In these applications, after raw data is transformed into a suitable format, it undergoes sophisticated query processing to derive high-level information. A related challenge is to capture uncertainty as data propagates through query operators until the final result and do so at stream speed.

In this paper, we present the design and development of a data stream system that captures data uncertainty from data collection to query processing to final result generation. To support uncertainty as a first-class citizen, our system models uncertain data items using *continuous random variables* (which are natural to most types of sensor data) and describes their uncertainty using probability density functions (pdf). The pdfs are transformed as data propagates through various operators. The final result can be described directly using its pdf or a confidence region, depending on the application. To capture uncertainty in high-volume stream processing, our system employs efficient techniques that are grounded in probability and statistical theory, and are particularly suited for manipulation of continuous random variables. Our system consists of two main components:

**1. Transforming raw streams into tuple streams with quantified uncertainty**. In real-world sensing applications, raw data streams emanate from devices without any uncertainty description. Transforming raw data by addressing both the mismatch between observed data and data of interest and the noise in observed data is a crucial initial step towards building an uncertainty-aware stream system. Recent research on sensor data cleaning and processing has applied probabilistic modeling and smoothing to temperature and light data [18, 19], GPS readings [35], and RFID data from static readers [21, 33]. However, these techniques do not address the mismatch of data formats in our target applications and performance requirements for such data transformation.

Our system takes a new approach to transforming raw data streams into tuple streams with quantified uncertainty. This approach fundamentally extends existing stream systems with the ability (*i*) to model the underlying data generation process and (*ii*) to use this model and principled probabilistic techniques to infer data of interest. More precisely, the inference procedure computes a distribution of values for the uncertain data needed in later processing, and outputs each inference result in a tuple carrying the obtained distribution. Our system performs such inference on high-volume streams by leveraging sampling and other advanced statistical techniques to achieve low cost and scalability.

**2. Relational processing of tuple streams under uncertainty.** As tuple streams propagate through various operators, our system also captures uncertainty of intermediate results as well as the final result. While the type of sensor data may vary in our applications, capturing uncertainty of processing results can be supported by a unified framework because data processing in these applications involves a common set of relational operators such as aggregation and join. To quantify result uncertainty of a relational operator, our system computes a distribution of values for each particular result that would be produced by the operator in a certain world. From this distribution, our system can also generate confidence regions and error bounds when needed.

There has been a recent surge of research on probabilistic databases [1, 3, 4, 9, 12, 38, 60, 62] and probabilistic stream processing [11, 31]. Most of existing work models tuples using *discrete* random variables and evaluates queries in a finite set of possible worlds. For efficiency, many systems avoid computing query result distributions by returning only a ranked list of results [48], decoupling and postponing result distribution calculation [3], or partially characterizing result distributions [11]. Unfortunately, these techniques do not apply to our problem for two reasons: (*i*) The continuous nature of sensor data dictates the use of *continuous* random variables and techniques suitable for them. (*ii*) Our goal to capture result uncertainty requires sufficient knowledge about the *entire* result distribution, which is especially important for uncertainty analysis of a pipeline of operators. This requirement precludes us from using existing techniques that compromise result distributions for simplified processing.

Among recent work on continuous random variables, Cheng et al. [9] supports aggregation of $n$ uncertain tuples using $n-1$ integrals, hence inefficient for stream processing. Other work [25, 30] generates samples over the distribution of $n$ random variables, runs query processing using these samples, and collects the results of these samples into a result distribution. It is an open question how many samples these algorithms need to capture complex real-world distributions and if they can scale for high-volume streams, e.g., 200Mb/sec.

Given our goal of accurate and fast characterization of each result distribution, we propose algorithms based on statistical theory that are natural to continuous random variables and particularly suited for their common distributions. It turns out that for many typical database operations such as aggregation and join over continuous variables, we can devise efficient algorithms for exact derivation of result distributions. If such results can be approximated with a small bounded error, we further consider approximation for even better performance. In certain complex scenarios where exact derivation is not possible, we explore fast, effective approximation suited for high-volume streams.

In the rest of the paper, we describe our target applications in more detail in Section 2, present our system architecture in Section 3, and outline technical issues in the two main system components in Sections 4 and 5.

## 2. MOTIVATING CASE STUDIES

In this section, we describe two sensing applications that our system aims to support.

### 2.1 Object Tracking and Monitoring

In the first application, RFID mobile readers are used to monitor a storage area such as a warehouse, a retail store, or a library. The storage area contains shelves $\mathcal{S}$ and objects $\mathcal{O}$ affixed with RFID tags. The shelf tags are at known locations but object tags are not. Usually, objects stay on the same shelf but sometimes move from one shelf to another. A mobile RFID reader provides the only means to observe the area. The reader can be either a handheld reader used by humans, or mounted on robots for automated monitoring and order processing (e.g., Kiva systems [37]). The mobile reader repeatedly scans the storage area. In each scan, it produces readings that contain the tag ids of observed objects, the tag ids of observed shelves, and optionally the location of the reader. These readings have two important characteristics, described as follows.

First, while the monitoring application wants precise object locations for further processing, the observed data simply contains observed tag ids—this is a fundamental limitation of an identification technology. Even if the reader returns its location, this does not directly reveal the object location, which can be twenty feet away in any direction. Second, the observed data is highly noisy. It is a well-known fact that the read rate of RFID readers is far less than 100% due to environment factors such as occluding metal objects, interference, and contention among tags. Moreover, mobile readers may read objects from arbitrary angles and distances, hence particularly susceptible to variable read rates.

Despite these data quality issues, the monitoring application needs accurate object locations to derive high-level information. We illustrate such needs using a fire monitoring application. Assume that raw RFID readings can be transformed into a stream of tuples each containing (`time`, `tag_id` of $O_i$, $(x, y, z)^p$), where $(x, y, z)^p$ denotes the uncertain $(x, y, z)$ location of the object.

The first query detects potential violations of a fire code:

display of solid merchandise shall not exceed 200 pounds per square foot of shelf area. The nested Select-From query simply adds two attributes to each tuple, the square foot area that each object belongs to, computed by a function on its $(x, y, z)^p$ location, and the weight of the object, retrieved by another function using its tag id. Then the outer query considers tuples in each 5 second window, groups them based on the square foot area, computes the total weight of the objects in each group. For groups with the total weight greater than 200 pounds, it reports the area and the total weight. The query is written as if the object location were precise, so the user does not need to worry about this issue.

```
Q1: Select   Rstream(R2.area, sum(R2.weight))
    From     ( Select Rstream(*, area(R.(x,y,z)) As area,
                              weight(R.tag_id) As weight)
               From   RFIDStream R [Now])
               R2 [Range 5 seconds]
    Group By R2.area
    Having   sum(R2.weight) > 200 pounds
```

The second query triggers an alert when a flammable object is detected in a area with a high temperature. The query takes two inputs: an object location stream as described above, and a temperature stream in form of (time, $(x, y, z)$, temp$^p$). The query selects flammable objects from one input and joins them with the temperature stream based on the equality of location. Again, the query is written as if the location of an object and the temperature at a location were precise.

```
Q2: Select   Rstream(R.tag_id, R.(x, y, z), T.temp)
    From     RFIDStream [Range 3 seconds] as R,
             TempStream [Range 3 seconds] as T
    Where    object_type(R.tag_id) = 'flammable' and
             T.temp > 60 ℃ and
             loc_equals(R.(x, y, z), T.(x,y,z))
```

## 2.2 Hazardous Weather Monitoring

The Engineering Research Center for Collaborative Adaptive Sensing of the Atmosphere (CASA) [7] is leading an effort to create distributed radar sensor networks with the goal to detect and monitor hazardous weather events like tornados [40]. To achieve this goal, CASA radar sensor networks are developed based on a closed loop system: In every 60 second cycle (called an epoch), data is transmitted from the radars to a central node where detection algorithms are executed. This information is then used to re-steer the radars, i.e., determining the scanning behavior of each single radar in the next epoch. This approach allows for the tracking of tornados or the scanning of a weather event with multiple radars to obtain a more accurate measurement.

The first CASA testbed consists of 4 radar nodes covering an area of 7,000 square km in southwestern Oklahoma and a central node that runs detection algorithms and controls the scanning of the individual radars. The region in which the testbed is located receives an average of four tornado warnings and 53 thunderstorm warnings each year [40].

A fundamental problem that emerges in this system is the possibility of detection errors caused by the uncertainty in the data generated by the radars and in the data transformed in various processing stages. In the following, data quality issues that can contribute to potential detection errors are described in more detail. The order of the description follows the path that the data takes through the system in each epoch,

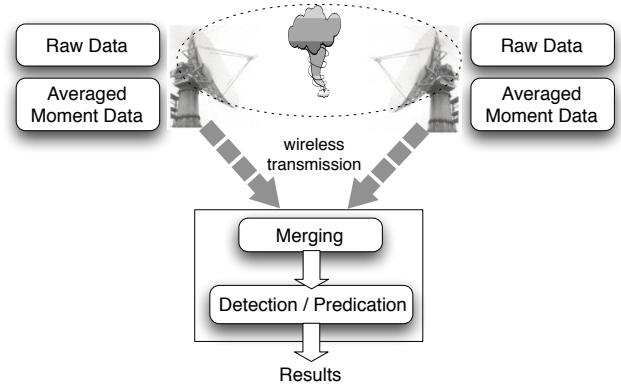

**Figure 1:** The path of data collection and processing in a CASA radar network for hazardous weather monitoring.

i.e., from the radars through wireless links to the center node as shown in Figure 1.

**High-volume raw data.** In each epoch, radars scan the atmosphere based on the control commands received from the central node. Each radar sends pulses at a rate of approximately 2000 pulses per second as it rotates to scan. A time series data structure is created for each pulse, which contains 832 range gates (measuring the distance from the radar) and a data item with four 32-bit floating numbers for each range gate. Hence, the raw data is generated at a rate of 1.66 million data items, or 205Mb, per second.

Noise in raw data can result from a variety of factors, including electronic device noise, instability of transmit frequency, quality issues of the system clock, the positioner, and the antenna, and finally environmental noise.

**Averaged moment data.** The radar's signal processor transforms the large amount of raw pulse data into moment data, which is a numeric description of each unit area of space (called a voxel) using real-valued attributes such as reflectivity, velocity, and spectral width.

Moment data can be generated for each data item in the raw pulse data. However, the resulting data volume may exceed the wireless bandwidth available for transmission to the central node and stress the tornado detection algorithm running there. Hence, in the current system moment data is generated by averaging over the data items from $N$ consecutive pulses but for the same gate. This averaging operation is essentially a temporal aggregation used to reduce data volume. But this operation may result in loss of precision in the processing of large-volume uncertain data.

**Merged data.** The central node merges streams of moment data from different radars that monitor overlapping regions. The merge procedure converts data from polar coordinates (centered at each radar) to Cartesian coordinates (latitude/longtitude/elevation), and fuses (or in the database terminology, joins) spatially overlapping data from multiple radars. Then meteorological detection algorithms are executed to detect tornados, severe storms, etc.

The data quality issue in the merging step is that the conversion from polar to Cartesian systems can cause uneven distribution of data density in the Cartesian system, hence affecting the quality of merged data. The issue is impaired by the fact that measurements for a certain Cartesian coordinate might be taken at different altitudes. For instance, the beams of two radars might overlap perfectly in the 2-dimensional plane but with a significant offset in the 3rd dimension (altitude).

**Table 1: Tornado detection using averaged moment data from 38 seconds of raw data taken on May $9^{th}$ 2007 during a tornadic event. The averaging size 40 is used to represent detection results using fine-grained data. The reported detection results are averaged over 4 sector scans in the 38 second period.**

| Averaging Size | Moment Data Size (MB) | Detection Running Time (sec.) | Num. of Reported Tornados | False Negatives |
|---|---|---|---|---|
| 40 | 9.22 | 27 | 3.75 | 0 |
| 60 | 6.15 | 23 | 1.5 | 2.25 |
| 80 | 4.62 | 21 | 0.5 | 3.25 |
| 100 | 3.7 | 21 | 0.25 | 3.75 |
| 200 | 1.87 | 20 | 0 | 3.75 |
| 500 | 0.76 | 20 | 0 | 3.75 |
| 1000 | 0.39 | 20 | 0 | 3.75 |

The quality issues mentioned above could lead to errors in the detection of meteorological hazards, including the failure to detect a tornado, misclassification of a detected tornado, and error in reported location of a detected tornado. Given the potential social impact of the CASA radar system, it is absolutely vital that it be augmented with the ability to capture data uncertainty and quantify result quality. This ability will enable end users, including National Weather Service forecasters, emergency managers, and scientists, to interpret detection results in a more meaningful way and to make better-informed decisions.

In our project, we aim to address all three uncertainty issues described above. We have performed an initial study to investigate the uncertainty regarding the averaging of moment data in more detail. In this study, we obtained 38 seconds of raw data taken in the CASA testbed on May $9^{th}$ 2007 during a tornadic event. We conducted an experiment in which the number of raw pulses used for averaging was varied. Table 1 shows the results for this experiment. Each row represents one run of the experiment where the only difference between experiments is the number of pulses averaged. To investigate the impact of averaging, we ran a tornado detection algorithm on the averaged data. Columns 2 and 3 of Table 1 show the file size of the averaged data and the running time of the detection algorithm (executed on a commodity PC with Intel Xeon 2.13 GHz CPU and 4GB of RAM). Columns 4 and 5 show the detection results, averaged over 4 sector scans of the radar in the 38 second period.

There are two constraints that have to be taken into account when evaluating the results of this experiment. First, as already mentioned the CASA system is a closed-loop system which has to operate on a 60 second cycle to be able to track fast moving weather events like tornados. In that 60 second period, there is roughly a 20 second time window allocated for the execution of the detection algorithm. Second, the available guaranteed bandwidth on the wireless links between the radars and the central node is 4 Mbps.

Taking these two constraints into account, it is evident from Table 1 that only the 500 and 1000 pulse averaging cases can be used in the existing system. The results also show that the uncertainty introduced by the high number of averaged pulses diminishes the quality of the results of the detection algorithm: in these two cases the detection algorithm would not report any tornado at all. This is a serious issue since there are spotter reports of at least a funnel cloud (which is basically a tornado but does not touch the ground) for the May $9^{th}$ event in the coverage area of the radar.

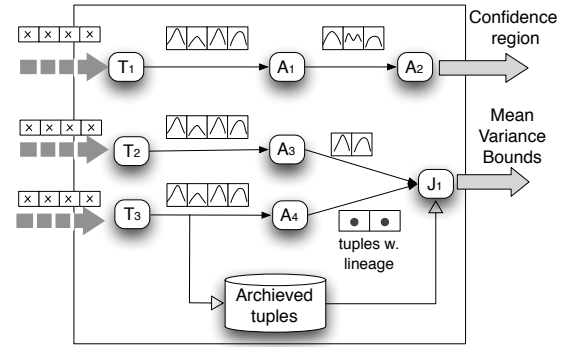

**Figure 2: Architecture of an uncertainty-aware stream system.**

On the other hand, using data with a less number of average pulses would increase the accuracy of the detection algorithm significantly. But the current CASA system does not have this information available because it cannot capture the effect of averaging over noisy radar streams. Once such information becomes available, the CASA system can decide dynamically to which data it can apply aggressive averaging without affecting the result, hence making CPU and bandwidth available for other data for which detailed analysis increases the quality of detection results significantly.

## 3. SYSTEM ARCHITECTURE

In this section, we describe the architectural design of an uncertainty-aware data stream system. Our system supports stream processing in the general box-arrow paradigm [6]. In this paradigm, a box represents a query operator and boxes are connected using arrows that represent the dataflow from one operator to another. This box-arrow diagram can be either compiled from a query (e.g., Q1 and Q2 in Section 2.1) or obtained from a scientific workflow (e.g., the workflow in the CASA radar system). Our system extends the box-arrow architecture in two aspects:

**Data Capture and Transformation (T) Operator.** The task of capturing uncertainty of raw data streams is encapsulated in a new "data capture and transformation" (T) operator. Allocated for each sensor device, a T operator serves as an ingress operator for the stream processing network. It offers two functions: First, it transforms raw data into a format suitable for further processing, e.g., a tuple stream with each tuple carrying an object location in the RFID application, or each tuple carrying velocity for each voxel in the radar application. Second, it includes a probability density function (*pdf*) in each output tuple. It is evident that to analyze uncertainty of further processing results, we need the pdf of each tuple—merely having mean, variance, and error bounds does not allow us to capture uncertainty of subsequent query processing results. Figure 2 illustrates such transformation for several T operators.

It is important to note that if the tuples produced by a T operator are independent, the pdf in each tuple completely characterizes its distribution. However, there are scenarios where the produced tuples are correlated, in particular, temporally correlated. In our system, the temporally correlated tuples, $X_1, X_2, ..., X_n$, each carry a conditional distribution $p(X_n|X_{n-1}, ..., X_{n-k})$ where $k \geq 1$. This way, a subsequent operator can construct their joint distribution, when needed, by multiplying these conditional distributions.

**Extended Query Operators.** Tuples output from the T operators are fed into downstream relational operators for further processing. Our system focuses on *selection*, *aggregation*, and *join* operators because they are common in query processing and particularly useful to our target applications. Besides their obvious application in the RFID example, they can also model various operations in the radar system. For instance, the averaging operation in moment data generation can be modeled using aggregation, and the merging of two radar streams is a special form of join.[1]

In our system, a query operator takes a sequence of tuples and produces a sequence of tuples that contain one of the following items:

- ▶ If the query operator is the last operator, its output tuples can carry full distributions, or alternatively, statistics such as the confidence region (a set of values whose confidence is over a threshold), mean, variance, or error bounds, depending on the application.
- ▶ If the query operator is an intermediate one and its output tuples are independent, each output tuple then carries its own distribution.
- ▶ If the query operator is an intermediate one and its output tuples can be correlated, each output tuple then contains its *lineage*, that is, a set of independent tuples produced from an upstream operator or the tuples from the initial T operators that were used to produce this tuple. If the input tuples of this operator are independent, it archives these input tuples for later computation of the query result distributions.

Figure 2 illustrates the third case using the operator A4. Its subsequent operator, also the final query operator, J1 uses the tuple lineage and previously archived independent tuples to compute its result distributions.

These two extensions of the existing data stream architecture are described in detail in the following sections.

## 4. UNCERTAINTY OF RAW DATA STREAMS

The foundation for building an uncertainty-aware stream system is to capture uncertainty of raw data close to the sensor that produces the data—this is a task of the data capture and transformation (T) operator in our system. To do so, we model the data in the transformed format as continuous random variables that cannot be directly observed (hidden variables $X$), and the data in the input format as continuous or discrete random variables that can be observed (evidence variables $O$). The task of data transformation and quantifying data uncertainty amounts to computing the conditional distribution $p(X|O)$.

---
[1]The final detection algorithm in the radar system is not relational. After knowing its input data quality, we will employ a suitable learning method to capture its result quality in our future work.

While our problem has traditionally been the purview of statistical machine learning [34], it poses a tremendous challenge to existing techniques due to the performance requirements of stream systems. For instance, our experimental results show that to transform a raw RFID data stream to an object location stream, a standard probabilistic inference technique can process only 0.1 reading per second for 20 objects [59]—this is neither efficient nor scalable for the data stream applications that we aim to support. Our work also differs from recent work in the database community that applies statistical or machine learning techniques to sensor types such as temperature and light [18, 19], GPS readings [35], and RFID data from static readers [21, 33]. This is because the new types of sensor data in our applications render different, and often more complex, statistical models, and yet pose more demanding performance requirements.

The main goal in our design of the data capture and transformation operator is to choose appropriate statistical machine learning techniques and further optimize them so that they can be performed for high-volume streams. Below, we describe a number of statistical techniques and optimizations employed in our system.

### 4.1 Modeling and Inference for Streams

The design of a data capture and transformation (T) operator consists of two steps: First, we model the underlying data generation process using a machine learning technique, calling *graphical modeling*, that captures how a sensor produces data from the true state of world with various types of noise. Next, we employ probabilistic inference to transform observed data into data of interest based on the data generation model. We illustrate this approach using mobile RFID data streams (more details are available in our recent publication [59]). The description in this subsection provides a technical context for the optimization issues discussed in the following subsections.

The first step in the design of a T operator is to develop a probabilistic model that captures the dynamic and noisy data generation process. Formally, the model is a joint probability distribution over all hidden and evidence variables in the problem domain. In the case of mobile RFID readers, hidden variables $X$ are the object locations, and evidence variables $O$ are the readings of objects, readings of shelf tags with known locations, and the reported reader location. Edges in the model capture dependencies among variables, e.g., the true object location and reader location jointly determining if the object is observed at time $t$.

Our graphical model is further divided into several components that describe how data is generated from the state of the world, e.g., the RFID sensing process, and how the state of world changes, e.g., the object location change. Each component is described using a local probability distribution. For instance, a distribution for RFID sensing can be devised using logistic regression over factors such as the distance and angle between the reader and an object. Then, the theory of graphical modeling allows us to write the complete joint distribution using the product of these local distributions.

The second step is to compute the distribution of hidden variables $X_t$ given observations $O_{1..t}$ from the joint distribution for the data generation process. In the RFID application, $p(X_t|O_{1..t})$ captures the distribution of true object locations given their observations by a mobile reader. The challenge is to perform accurate inference at stream-speed and for a large number of objects.

We approach this problem by adopting sampling-based inference since exact inference is intractable for our problem. A common method, called *particle filtering*, maintains a list of samples (termed particles) of the state of all hidden variables and weights the samples based on all observations received thus far. Over time the weighted particles become an approximation of the target conditional distribution.

Our system employs a series of optimizations of a particle filter to achieve efficiency and scalability in stream processing. *Factorization* breaks a large particle over all hidden variables into smaller particles over individual hidden variables while obeying the dependency constraints in our graphical model. This allows us to reduce the worst case of an exponential number of particles to a linear number of particles for each variable (object). After factorization, *spatial indexing* can further limit the set of variables that must be processed at each time step, since a reader can only observe a small set of objects at a time. After object particles stabilize in a small region, *compression* can further reduce the number of particles needed for each object. Through these optimizations, our system improves particle filtering from processing 0.1 reading per second given 20 objects to over 1000 readings per second in most cases given 20,000 objects, e.g., achieving 7 orders of magnitude improvement in scalability.

## 4.2 Exploring Speed-Accuracy Tradeoffs

Sampling-based inference presents a fundamental tradeoff between accuracy and performance. Figure 3(a) and 3(b) show the results of performing inference for a highly noisy trace of RFID readings. In particular, these figures show that the inference error reduces but the computation cost increases as we increase the number of particles used for each object. Hence, it is important to determine this tradeoff dynamically given application requirements.

If the application has an accuracy requirement, our system finds the minimum number of particles that allows inference to meet the accuracy requirement. To measure inference accuracy dynamically, our system uses reference objects with known true information. In the RFID example, these objects are shelf tags at fixed locations that are known a priori. Then our system (conceptually) replicates the nodes of these objects in the graphical model such that one copy serves as evidence variables and the other as hidden variables. This way, we can obtain estimated locations from the copy of hidden variables and compare them with the true locations to measure inference accuracy. To dynamically adjust the number of particles, our system uses a feedback control method: it starts with a relatively small number of particles and keeps doubling this number before meeting the accuracy requirement. After that, it reduces the number of particles by a constant each time until it finds the smallest number. A similar method can be used to maximize accuracy while meeting the application performance requirement.

## 4.3 Approximating Result Distributions

The third issue is how to generate an output stream with sufficient statistics about uncertainty. In the RFID example, each tuple in the output stream describes the estimated location of an object. Some applications may require only the confidence region of the estimated location, e.g., with 90% confidence that the object is in a certain range. Some other applications, however, may require further processing of the location stream. To capture uncertainty of such further processing, each tuple in the location stream must carry the full distribution of the object location. We call the distribution in each tuple the *tuple-level distribution*.

A direct way to generate a tuple-level distribution is to include in each tuple the weighted samples (particles) used in inference. This is naturally a sample-based representation of the tuple-level distribution. An obvious problem with this approach is that every tuple now carries tens or hundreds of samples. This will increase the stream volume by one or two orders of magnitude. In addition, it will slow down further query processing because those samples need to be processed one at a time.

For both time and space efficiency, our system converts a sample-based tuple level distribution into an approximate *parametric* distribution such as a Gaussian distribution or more flexible distributions. Our system does so by minimizing the KL divergence (a standard measure of "distance" between distributions) $\text{KL}(\hat{p}\|q)$, where $\hat{p}$ is the sample-based tuple level distribution, and $q$ is the target parametric distribution. Assume $\hat{p}$ to be a list of value-weight pairs, $\{(x_i, w_i)\}$. Consider a Gaussian distribution $N(\mu, \sigma^2)$ for $q$. Then,

$$\text{KL}(\hat{p}\|q) = \sum_i w_i \cdot log \frac{w_i}{q(x_i)}$$
$$= \sum_i w_i \cdot log(w_i \cdot \sigma \cdot \sqrt{2\pi}) + \sum_i w_i \cdot \frac{(x_i - \mu)^2}{2\sigma^2}$$

Minimizing $\text{KL}(\hat{p}\|q)$ allows us to find the optimal Gaussian parameters to represent $\hat{p}$. That is, $\mu = \sum_i w_i \cdot x_i$ and $\sigma^2 = \sum_i w_i \cdot (x_i - \mu)^2$. Hence, we can efficiently convert a sample-based distribution to the closest Gaussian using two scans of the list of samples. Similar formulas are available to convert sample-based distributions into multivariate Gaussians.

While approximating a sample-based distribution using a Gaussian distribution is efficient, there are scenarios where we have to consider more flexible parametric distributions to reduce the error of such approximation. In the RFID application, an object may have recently moved from one location to another. The samples for this object's location can be temporarily spread over two locations. Approximating these samples using a single Gaussian is obviously inaccurate. Then a mixture of Gaussians may be appropriate, in which one component of the Gaussian corresponds to the possibility that the object has not moved, while the other represents possible new locations of the object. Selecting the number of mixture components, that is, the number of "humps" in the mixture, can be done using standard model selection techniques such as Akaike Information Criterion (AIC) and the Bayesian Information Criterion (BIC). Both of these techniques attempt to choose a number of components that explain the data well while penalizing models that require many mixture components.

## 4.4 Alternative Techniques for Extremely High Volume Streams

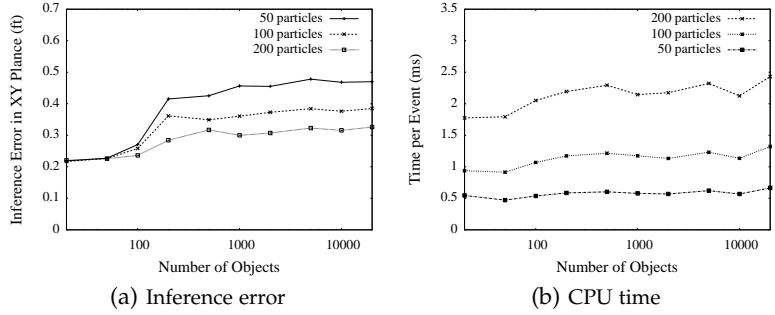

(a) Inference error  (b) CPU time

**Figure 3: Accuracy and performance results for a high noisy RFID trace.**

We described above the design issues for the data capture and transformation operator using the RFID application. We are also investigating these issues for the weather monitoring application. In the second application, the raw data stream contains pulse data from each radar and the transformed stream contains a tuple for each voxel that has moment data including reflectivity, velocity, etc. for that voxel.

Our idea of modeling the data generation process and augmenting the transformed stream with a distribution for each tuple still applies. However, this application presents two challenges to graphical modeling that aims to completely characterize the data generation process. The first challenge is the complexity of the data generation process. As mentioned in §2.2, the quality of observed data is affected by many factors such as environmental noise, electronic device noise, instability of transmit frequency, and inaccuracy of the system clock, the positioner, and the antenna. Precisely describing a data generation process involving all these factors requires significant domain knowledge. The second challenge is that the data volume in this application is extremely high, e.g., 1.66 million data items (205Mb) per second. It is an open question if sampling-based inference, even with optimizations, can keep up with such stream speeds.

To cope with the complexity of the problem and the extremely high stream volume, we seek alternative modeling techniques that allow us to quickly obtain an approximation of the distributions of the moment data. We observe that in this application, the transformation from the raw data to the moment data is deterministic and based on precise formulas (unlike the RFID application). Hence, we can obtain the transformed moment data stream and characterize its uncertainty using a relatively simple time series model. For a concrete example, consider the velocity data for a particular voxel. We denote the obtained velocity series using variables $O_1, ..., O_t$, and their true values using variables $X_1, ..., X_t$. We can describe the uncertainty of velocity data using $p(X_1, ..., X_t | O_1, ..., O_t)$. To do so, we consider the Autoregressive Moving Average (ARMA) Model that captures how $X_t$ relates to the previous variables (autoregression) and the noise factors in the recent period (moving average). Formally,

$$X_t = \sum_{i=1}^{p} a_i X_{t-i} + \sum_{i=1}^{q} b_i \epsilon_{t-i} + \epsilon_t + C$$

where $\epsilon_t$ is the noise term for time $t$, and $C$ is a constant. There are well known numeric methods that given observed data, find the ARMA($p,q$) model together with the coefficients that best fits the data. These fitting methods, however, may take many passes over the data to find the best fitting. Hence, they may still be slow for the stream volume in the CASA system.

For improved efficiency, we reduce the overhead of modeling to the minimum using two techniques. First, we model $X_t$ simply using the moving average model (MA) with no autoregression. This assumption is likely to hold for a short sequence of data: due to frequent sampling, a short sequence of data tends to describe the same phenomena, hence obviating the need of autoregression, but with correlated noise factors. As such, the work needed for modeling is to identify sequences where the MA assumption holds. Based on statistical theory, sequences obeying the MA assumption can be identified by computing their $k$-lag autocorrelations, which can be performed using at most two scans of the input sequence. Second, if we know that the first query operator on the transformed stream is aggregation such as average and sum, which is true in the CASA system, we do not need to fit the MA model precisely. This is because we can use the Central Limit Theorem to obtain asymptotic results for aggregation, disregarding the precise input distributions, as long as the MA assumption holds.

## 5. UNCERTAINTY OF QUERY PROCESSING RESULTS

After uncertain tuple streams are generated from the raw data from each sensor device, they then go through various operators to produce final results. Hence, it is also important to capture uncertainty of such processing results. Our system considers common relational operators, including selection, aggregation, and join, which provide general support for both of our target applications. Our system quantifies result uncertainty of each query operator by computing a distribution for each result tuple. More precisely, a result distribution is a distribution of values for each tuple that would be produced by the operator in a certain world, e.g., the total weight for each square foot area for Q1 and the temperature of each object for Q2 from Section 2.1. The main goal is to characterize result distributions of these operators accurately and efficiently for high-volume streams.

Characterizing result distributions for inputs involving continuous random variables is a difficult task. Existing solutions have been based on either an integral-based approach [9] or a sampling-based approach [25, 57]. The former approach is an exact derivation of the result distribution, based on statistical theory for continuous random variables. The derived distribution is accurate but the computation is slow. For instance, for aggregation over *n* input tuples, the two variable convolution algorithm in [9] uses *n*-1 integrals. The latter approach discretizes the continuous distributions and samples from the discretized distributions. When a small number of samples is used, the computation is fast but with potential loss of accuracy. To accurately characterize complex real-world distributions, the number of samples needed can be large, and hence the performance suffers.

A main idea underlying our work is to explore advanced statistical techniques that are natural and particularly suited for continuous random variables. These techniques include using common continuous distributions and Gaussian mixture distributions as models, and using characteristic functions and order statistics to compute result distributions directly. These techniques allow us to perform exact derivation of result distributions without using integration or using it to the minimum extent. Furthermore, our work applies approximation whenever possible to simplify computation. Such approximation techniques have the promise of achieving a balance between speed and accuracy in a more effective way than sampling-based methods.

Below, we illustrate our ideas using aggregation operators (but leave the detailed techniques for aggregation and other operators to our future work). We also discuss issues related to composed operators at the end of the section.

## 5.1 Aggregation Operators

An aggregation operator takes *N* tuples, modeled as *N* random variables, and performs an operation, such as sum or max, on these variables. The correlation structure among these variables determines appropriate techniques. Domain knowledge can be used to infer the correlation structure. In the RFID example, object locations can be considered independent if domain knowledge reveals that objects move in space independently. In the radar system, however, the data items for the 2000 pulses in each second form a correlated time series, due to frequent sampling. Given a realized sequence of *N* random variables, model testing and identification tools ([5], Chapter 9) can be used to test the randomness and determine the order of correlation if it does exist.

**Independent variables**. We take sum as an example in the following discussion. To sum *N* independent variables, the exact result distribution can be obtained through inversion of the characteristic function (CF) of the sum, which is the product of the characteristic functions of the individual summands. For variables with common distributions, the characteristic functions usually have closed form representations, which greatly reduces the computation cost. This way, the inversion expresses the exact result distribution using a single integral, in contrast to *N*-1 integrals in [9]. If the single integral is still time-consuming to evaluate, we can further approximate the result distribution using Gaussian or mixture of Gaussian distributions. The parameters of these distributions can be identified by fitting the characteristic functions of the Gaussian or mixture of Gaussian distributions to the closed form characteristic function of the sum.

To investigate these ideas, we performed an initial experiment that compares the performances of the histogram-based sampling algorithm [25], the inversion of the characteristic function, and the approximation of the characteristic function. The input distributions are different for different tuples, and are generated from mixture Gaussian distributions to simulate arbitrary real-world distributions. We use the exact result distribution calculated from the inversion of the characteristic function as a criterion to calibrate the accuracy. We compute the variance distance to capture the distance between the exact distribution and the output of each of the other two algorithms, based on the formula in [25]. Table 2 summarizes the results of performing sum over non-overlapping windows of *N*=100 tuples.

**Table 2: Algorithm comparison for performing sum over a tuple stream. A tumbling window of size of 100 tuples is used for aggregation.**

| Algorithm | Throughput | Variance Distance∈ [0,1] |
|---|---|---|
| Histogram | 3382 | 0.083 |
| CF (inversion) | 466 | 0 |
| CF (approx.) | 10593 | 0.012 |

These results show that for the given workload, the approximation of the characteristic function performs the best in terms of both speed and accuracy. The inversion-based method gives the exact result distribution, but is too slow for stream processing as the result of using a single integral. (Hence we conclude that the algorithm using *N*-1 integrals in [9] is infeasible for stream processing). The histogram-based sampling method does not achieve a good balance between speed and accuracy.

Additional approximation techniques are available. For instance, approximation to the result distribution can be obtained through the Central Limit Theorem for sum of independent variables, when the number of the effective summands is fairly large ([54], Chapter 1.9). When certain conditions are met, the Central Limit Theorem states that the distribution of the sum converges to a normal distribution, regardless of the distributions of the summands. When this can be applied, the compution cost for the result distribution is almost zero.

**Correlated variables**. If the input tuples are correlated, e.g., forming a time series, exact derivation of the result distribution of sum can be difficult, although not impossible. For the given joint distribution of the summands, the distribution of the sum can be obtained through multivariate integrations. Numerical methods such as adaptive quadrature or Monte Carlo integration can be used [42].

Approximation techniques also depend on specific correlation structures. One technique that pertains to our hazardous weather monitoring is the Central Limit Theorem for time series [6]. As stated in section 4.4, in the observed velocity series, we can use the autocorrelation function to identify sub-series length to aggregate upon so that the subseries can be modeled as an MA model. For a series that is from an MA model, the Central Limit Theorem states that the average velocity has an asymptotic normal distribution, of which the mean and variance can be estimated based on the sample mean and sample autocorrelation function.

## 5.2 Composed Operators

If a query involves multiple operators, the intermediate results can be correlated even if the input tuples to the first operator(s) are independent. We need to carefully examine when such correlation occurs. In the radar system, for instance, the raw data first undergoes temporal averaging across consecutive pulses and then a merge (join) between streams from different radar nodes. The first average, however, does not create correlated results because it is applied to non-overlapping segments of the raw data stream—such aggregation is very common in scientific applications (also reported in [29]). On the other hand, if a join is followed by an aggregation, the join may produce correlated results by matching a tuple from one input with multiple tuples from the other. Then characterizing result distributions of aggregation with correlated inputs requires the full joint distribution of input tuples and is hard to compute.

A general solution to computing result distributions from correlated intermediate tuples is to use sampling and density estimation to obtain the result distribution. However, this can be slow for high-volume streams. Given our focus on selection, join, and aggregation and their uses in real-world applications, we aim to identify common patterns of correlation and explore several types of optimization and approximation to obtain the result distributions.

**Complex functions**. A simpler solution is possible if we can treat several composed operators as a single complex function. Then we only need to deal with independent inputs. Such complex functions are possible for a series of mathematical computations (e.g., aggregation). Even with join and aggregation combined, in some cases it is possible to write the combined operation using a single continuous function. Take Q2 from section 2.1 as the join example. One way to do the join is to first smooth the temperature trace and express the temperature as a function $h$ over the inferred location of an object. Now if we were to aggregate temperature over a set of objects, the aggregation result can be expressed using a complex function over a set of temperature functions, one for each object. If the complex function is differentiable, we can apply the transformation theory for continuous random variables to obtain the exact result distribution, or the multivariate Delta method to approximate the result distribution for efficiency.

**Lineage.** Another useful technique is to exploit lineage [3, 62] about how correlated tuples are produced. Such lineage helps determine the correlation structure among tuples and eliminates the need of computing full joint distributions for intermediate tuples. Given the correlation structure, the last query operator has the flexibility to optimize its computation. For instance, the last operator can use fast techniques for the subset of independent tuples and more sophisticated techniques for the subset of correlated tuples. When this operator computes result distributions for a set of tuples with overlapping lineage structures, it can apply optimizations to compute for all these tuples in a shared manner (similar to [52]). Furthermore, it may also be possible to find approximate lineage [50] that gives a good approximation of the result distributions and allows more efficient computation.

In our immediate future research, we will explore a number of issues related to applying lineage to uncertain data streams where tuples are modeled using continuous random variables. Given our focus on selection, join, and aggregation, we will study how to compute result distributions using lineage for this set of operators. In particular, aggregation significantly complicates the computation and hence remains unaddressed by existing work on modeling uncertain data using continuous variables [57]. We will also address the efficiency of computation and optimizations mentioned above, which have been mostly studied for probabilistic databases, in the new context of stream systems. We will further consider compact representations of lineage to reduce the volume of intermediate streams passed between operators.

## 6. RELATED WORK

**Sensor data management** has been an area of intensive recent research [43, 64, 44, 18, 10, 16, 19, 63, 55, 56, 65]. The sensor data considered in these studies contains incomplete and noisy measurements of environmental phenomena such as temperature and light. Regarding modeling, the work most relevant to ours [18, 17, 16] builds statistical models to capture correlations among attributes and attribute value changes, enabling reduced sensing rates while meeting a query-specified confidence. FunctionDB [58] transforms discrete sensor observations into continuous functions and supports querying over these functions.

Our work differs in three aspects. First, we consider different types of sensor data, i.e., RFID data from mobile readers and meteorological data from radar networks. Our sensing processes are different and more complex, hence requiring more sophisticated modeling techniques. Second, our work also quantifies uncertainty of query results as noisy data goes through various processing stages, which has been under-addressed in existing sensor data research. Third, our work captures data uncertainty over high-volume streams and hence employs many optimizations for efficiency and scalability in such stream processing.

**Sensor data stream cleaning** aims to build an abstraction of device data appropriate for query processing [21, 32, 61], which is similar to our work on capturing uncertainty of raw sensor data. SMURF [33] considers RFID readings generated by static readers and mitigates the effect of missed readings by applying temporal and spatial smoothing. SMURF and other recent studies on RFID stream cleaning [61, 49] offer coarse-grained location information, e.g., whether an object is in the large read range of the reader or in an office, which may not be precise enough for object tracking in our target application. In addition, these studies consider at most one hundred objects. Probabilistic inference techniques such as Kalman filters have also been used to filter and smooth noisy GPS readings [35]. In comparison, our work employs different modeling techniques to handle mobile RFID data and radar data, and explores optimizations to do so on high-volume streams. Our work also captures the quality of query processing results from such data.

**Probabilistic databases** have been an area of intensive recent research[1, 3, 4, 9, 12, 38, 36, 53, 60, 62]. In a probabilistic database, each tuple exists with a certain probability [2, 8, 12, 41]; such tuple existence is essentially characterized by a Bernoulli distribution. Each tuple may further contain a distribution over a set of values [51]. Given such tuple models in a *discrete and finite* domain, a probabilistic database is a probability distribution over all database instances called *possible worlds* [12]. Query evaluation applies a query to each possible world, and adds the probabilities of all possible worlds that return the same answer, yeilding a distribution of possible query answers. Dalvi and Suciu [12, 13] have identified cases when one can compute the result distribution directly from the probababilities of base tuples and when one has to consider all possible worlds. For efficiency, several studies attempt to avoid the computation of the result distribution by simply returning a ranked list of results [48] or using lineage to decouple and postpone the computation of result probabilities [3, 46, 52, 62].

There have been a few recent studies that model uncertain data using continuous random variables [9, 25, 30, 57]. Among them, Cheng et al. [9] support SUM and COUNT over $n$ uncertain tuples by integrating two variables at a time, resulting in a total of $(n-1)$ integrals. Since integrals are very expensive to compute and the number of tuples for a single aggregation can be large, this algorithm is infeasible for stream processing. Other work [25, 30] generates samples over the distribution of $n$ random variables, runs query processing using these samples, and collects the results of these samples into a result distribution. A concern with these algorithms is the need of a large number of samples to achieve accuracy for arbitrary real-world distributions, hence slow for high-volume streams.

Our work differs from probabilistic databases in several important aspects. First, we capture certainty of both raw data and results of processing such data, whereas probabilistic databases assume that probabilities of tuples already exist. Second, due to the continuous nature of sensor data, our work employs techniques that are natural and suitable for continuous random variables and are fundamentally different from those for discrete variables based on the possible world semantics. Finally, we capture data and query result uncertainty over streams, in contrast to doing so within a database, and thus focus on time and space efficiency of our techniques.

**Probabilistic stream processing** has gained research attention very recently. Issues in such processing over sensor data streams are discussed in [23]. Existing work [11, 31] adopts the finite and discrete tuple model used in probabilistic databases. As stated above, many of the techniques for discrete variables cannot be applied for problems involving continuous variables. Furthermore, the existing work produces mean, variance, and higher moments to describe the result distributions, which cannot be easily used to compute the distribution of subsequent operators.

**Approximate query processing** [15, 24, 26, 27, 28, 14, 45, 47, 10] addresses a fundamentally different problem: While the uncertainty of raw data and query results computed from such data is inherent in our problem, approximate query processing trades off exact query processing for approximate answers to improve query performance, to adapt to resource constraints, or to save energy consumption.

## 7. SUMMARY

As of December 2008, we have implemented a prototype system that includes a data capture and transformation (T) operator for mobile RFID data streams, and an initial simple T operator for radar data streams. We are currently building a library of techniques for efficiently deriving result distributions for selection, join, and aggregation. We have collected traces from both of our target applications. We are evaluating individual techniques using these traces and plan to integrate them into a full stream system using an existing data stream implementation such as [29]. We hope to eventually evaluate our techniques in the operational CASA radar system.

## 8. ACKNOWLEDGMENTS

This work has been supported in part by the National Science Foundation under the grants IIS-0746939 and IIS-0812347, and EEC-0313747. We would like to thank Prashant Shenoy, Richard Cocci, and Yanming Nie for their contributions to the RFID case study. We also thank Ming Li, David Pepyne, and Eric Lyons for their valuable help with the CASA case study.